\documentclass{PoS}
\newcommand{\be}{\begin{equation}}
\newcommand{\ee}{\end{equation}}
\newcommand{\beq}{\begin{eqnarray}}
\newcommand{\eeq}{\end{eqnarray}}
\newcommand{\bea}{\begin{eqnarray*}}
\newcommand{\eea}{\end{eqnarray*}}

\usepackage{epsfig}
\usepackage{epsf}
 \usepackage{dsfont}
\usepackage{slashed}
\usepackage{booktabs}
\usepackage{graphicx}
\usepackage{float}

\title{Nucleon Structure in Lattice QCD using twisted mass fermions}

\ShortTitle{Nucleon Structure in Lattice QCD}

\author{\speaker{C. Alexandrou}~$^{(a,b)}$, M. Constantinou~$^{(a)}$, T Korzec~$^{(a,c)}$,\\
$^{(a)}$ Department of Physics, University of Cyprus, P.O. Box 20537, 1678 Nicosia, Cyprus\\
$^{(b)}$ Computation-based Science and Technology Research
    Center, Cyprus Institute, 20 Kavafi Str., Nicosia 2121, Cyprus    \\
$^{(c)}$ Institut f\"ur Physik
   Humboldt Universit\"at zu Berlin, Newtonstrasse 15, 12489 Berlin, Germany\\
        E-mail: \email{alexand@ucy.ac.cy}, \email{constantinou.martha@ucy.ac.cy}, \email{korzec@physik.hu-berlin.de}}
\author{J.~Carbonell,  P.~A.~Harraud, M.~Papinutto\\
 Laboratoire de Physique Subatomique et Cosmologie,
               UJF/CNRS/IN2P3, 53 avenue des Martyrs, 38026 Grenoble, France\\
E-mail:\email{Jaume.Carbonell@lpsc.in2p3.fr}, \email{harraud@lpsc.in2p3.fr}, \email{Mauro.Papinutto@lpsc.in2p3.fr}}
\author{P. Guichon\\
CEA-Saclay, IRFU/SPhN, 91191 Gif-sur-Yvette, France\\
        E-mail: \email{pierre.guichon@cea.fr}}
\author{K. Jansen\\
NIC, DESY, Platanenallee 6, D-15738 Zeuthen, Germany\\
        E-mail: \email{Karl.Jansen@desy.de}}

\abstract{We present results on the nucleon form factors and moments of 
generalized parton distributions obtained within the twisted mass 
formulation of lattice QCD. We include a discussion of lattice artifacts
by examining results at different volumes and lattice spacings. We compare
our  results with those obtained using different discretization schemes and to
 experiment. }

\FullConference{35th International Conference of High Energy Physics - ICHEP2010,\\
		July 22-28, 2010\\
		Paris France}

\begin{document}

\section{Introduction}
Lattice QCD simulations are currently being performed  with two dynamical  degenerate light quarks with a mass close  to their physical value 
as well as  the strange quark using a number of different
discretization schemes with the most common being 
Wilson-improved, staggered and chiral fermions.
Furthermore, simulations at several lattice spacings and volumes are 
becoming available enabling a comprehensive study of lattice artifacts.

The focus of this contribution is the evaluation of form factors (FFs) and  moments of parton distributions of the nucleon, which are being measured in many experiments.
The characterization of nucleon structure is considered a milestone in hadronic physics and experiments on nucleon FFs started in the 50s. A new generation of experiments using polarized beams and targets are yielding high precision data spanning a larger range of momentum transfers. FFs provide ideal probes of the charge and magnetization densities of the hadron as well as a determination of its shape.

 Lattice techniques to extract nucleon matrix elements connected to FFs and moments of generalized parton distributions (GPDs) are well developed.
 The connected diagram where on operator couples to a valence quark  is 
straightforward to compute and has been evaluated by a number
of lattice groups~\cite{Alexandrou2010}.
For iso-vector nucleon matrix elements, in the isospin limit, 
this is the only contribution. Like most collaborations, 
we use non-perturbative renormalization of these matrix elements and furthermore we subtract ${\cal O}(a^2)$-terms computed perturbatively to improve the extraction of the renormalization constants~\cite{Z-Alexandrou}. 
Using simulations of $N_F=2$ twisted mass fermions (TMF) at three
values of the lattice spacing and different volumes we
study cut-off and volume effects. We use the lattice spacing determined
from the nucleon mass to convert to physical units.
Heavy baryon chiral perturbation
theory (HB$\chi$PT) is used to  extrapolate lattice results obtained for pion masses in the range of about 260~MeV to 470~MeV to the
physical point.
 
 \vspace*{-0.3cm}

\section{Nucleon form factors} \vspace*{-0.3cm}
The nucleon matrix element of the electromagnetic (EM) current,
$j_\mu=\bar{\psi}(x)\gamma_\mu \psi(x)$,  is written in the form
  $\bar{u}_N(p^\prime,s^\prime)\left[\gamma_\mu { F_1(q^2)} + \frac{i \sigma_{\mu\nu} q^\nu}{2 m} F_2(q^2)\right] u_N(p,s)$,
where the Dirac {$F_1$} and Pauli {$F_2$}  FFs are  related to the electric and magnetic Sachs FFs with the relations:
  ${G_E(q^2)} = {F_1(q^2)} - \frac{q^2}{(2m)^2}{F_2(q^2)} $ and
     ${G_M(q^2)} = {F_1(q^2)} + {F_2(q^2)}$.
For the axial vector current, $A^a_\mu=\bar{\psi}(x)\gamma_\mu\gamma_5 \frac{\tau^a}{2}\psi(x)$,  the nucleon matrix element is of the form
$ \bar{u}_N(p^\prime,s^\prime)\left[ \gamma_\mu\gamma_5 { G_A(q^2)} + \frac{q^\mu\gamma_5}{2 m} {G_p(q^2)} \right] \frac{1}{2} u_N(p,s)$.

\begin{figure}[h]\vspace*{-4.5cm}
\begin{minipage}{0.55\linewidth}
  { \hspace*{-1cm}\includegraphics[width=1.1\linewidth]{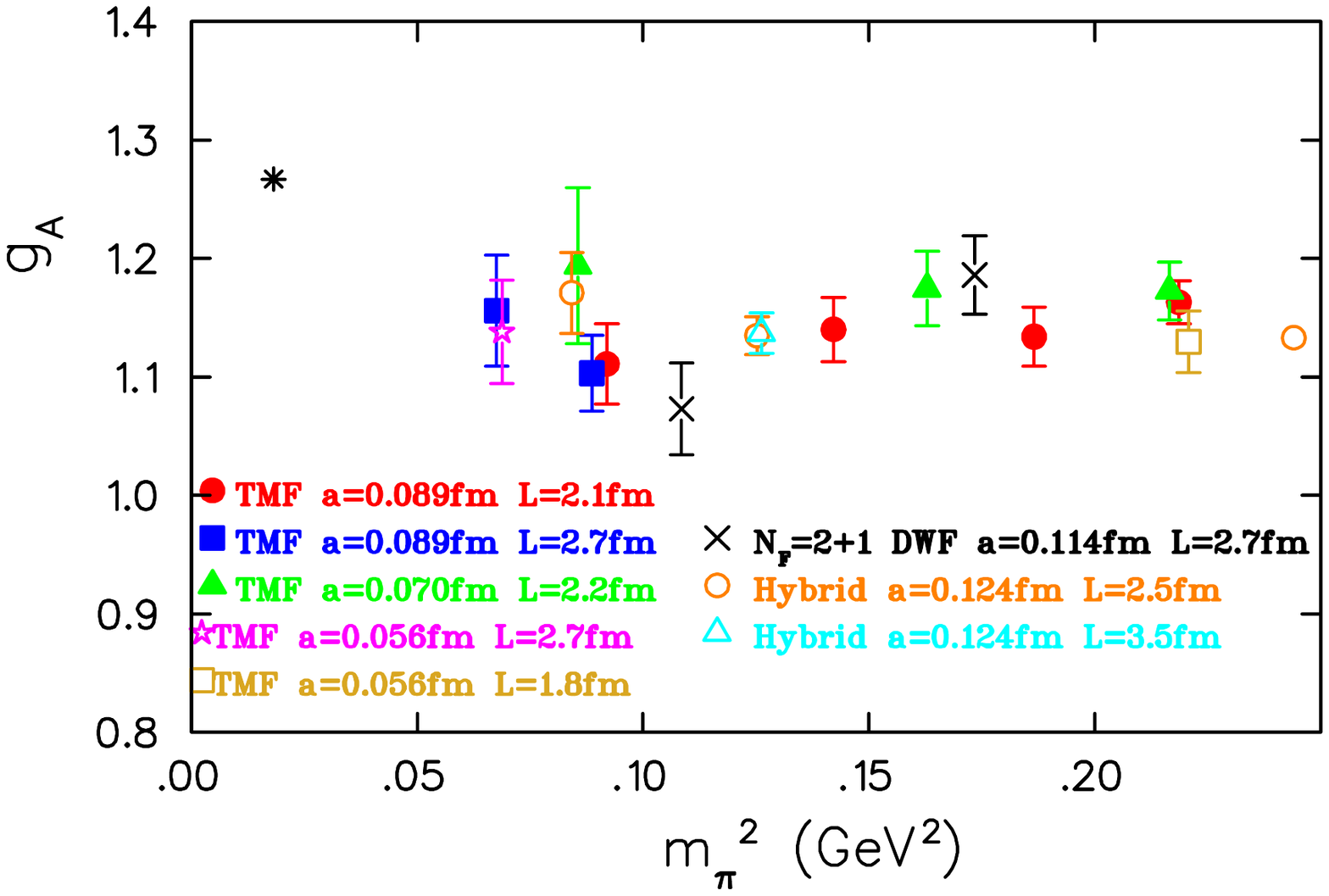}}
\end{minipage}\hfill
\begin{minipage}{0.55\linewidth}
  { \hspace*{-1cm}\includegraphics[width=1.1\linewidth]{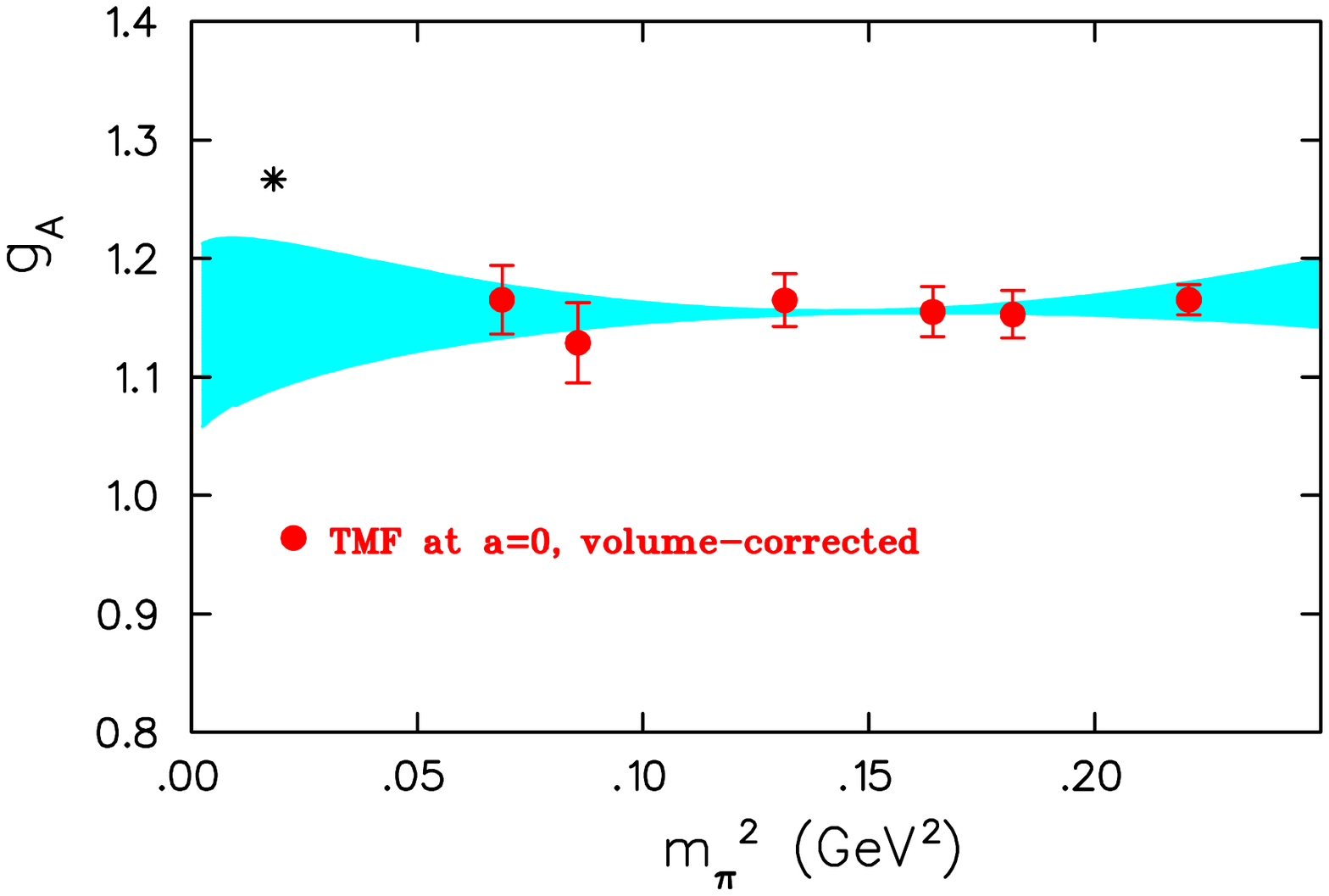}}
\end{minipage}\vspace*{-2.5cm}
\caption{Left: Lattice data on $g_A$ using $N_F=2$ TMF~\cite{Alexandrou2010},  $N_F=2+1$  DWF~\cite{FF-Yamazaki} and $N_F=2+1$ using DWF on staggered sea quarks~\cite{FF-Bratt}. The physical point is shown by the asterisk.
Right: Volume corrected continuum TMF  results  together with the band obtained using HB$\chi$PT.} 
\label{fig:gA}
\end{figure}

The axial charge is well known experimentally. Since it is determined at $Q^2=0$
there is no ambiguity associated with fitting the $Q^2$-dependence
 of the FF. As can be seen in Fig.~\ref{fig:gA}, where we show  recent
 lattice results using TMF, domain wall fermions (DWF) and a hybrid action of DWF valence on staggered sea quarks, there is a nice 
 agreement among different 
lattice discretizations and no significant dependence on the quark mass down
to about $m_\pi=270$~MeV.

\begin{figure*}[h]\vspace*{-3.5cm}
\begin{minipage}{0.49\linewidth}
      {\includegraphics[width=1.1\linewidth]{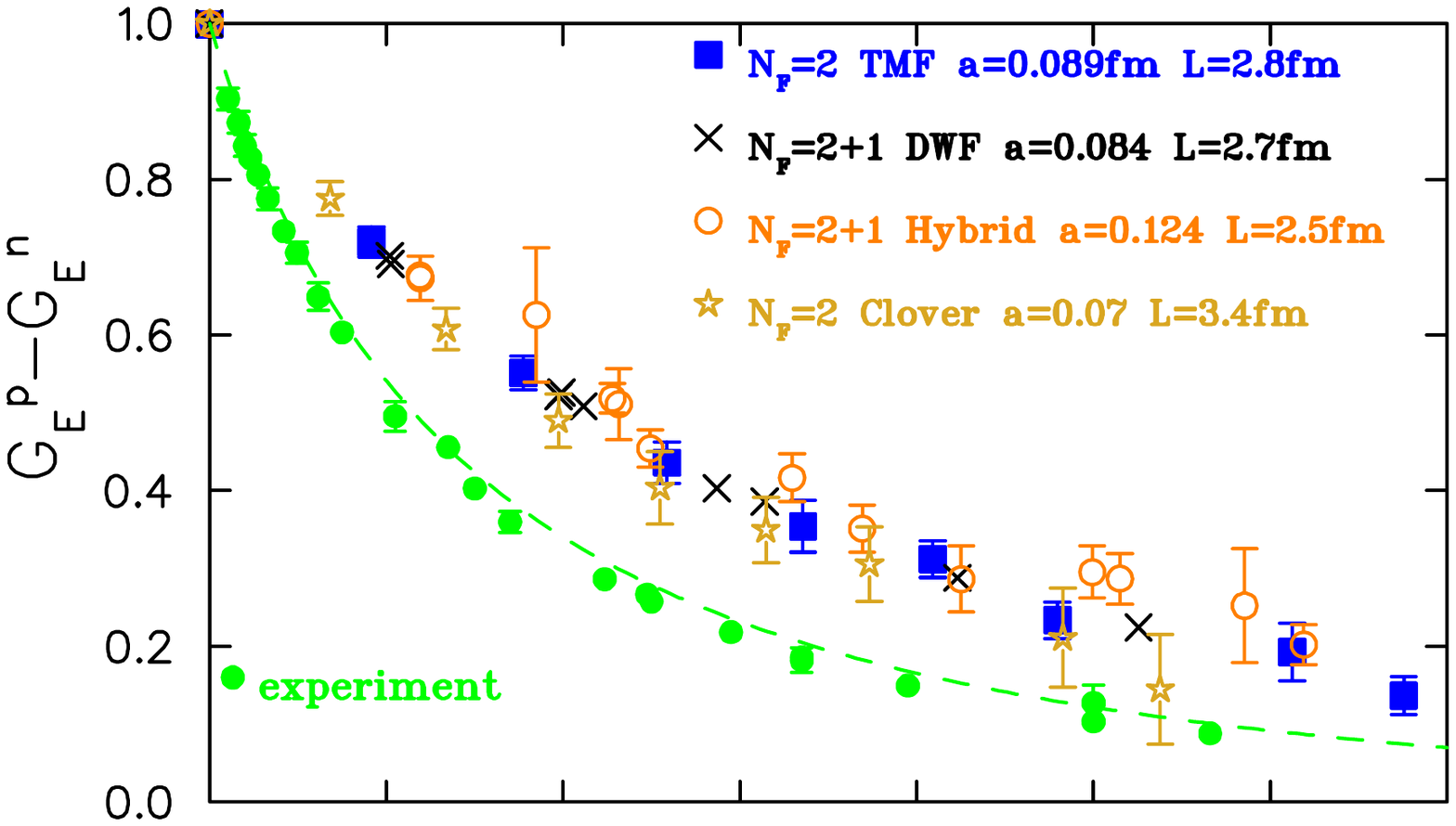}}\vspace*{-6.8cm}
      {\includegraphics[width=1.1\linewidth]{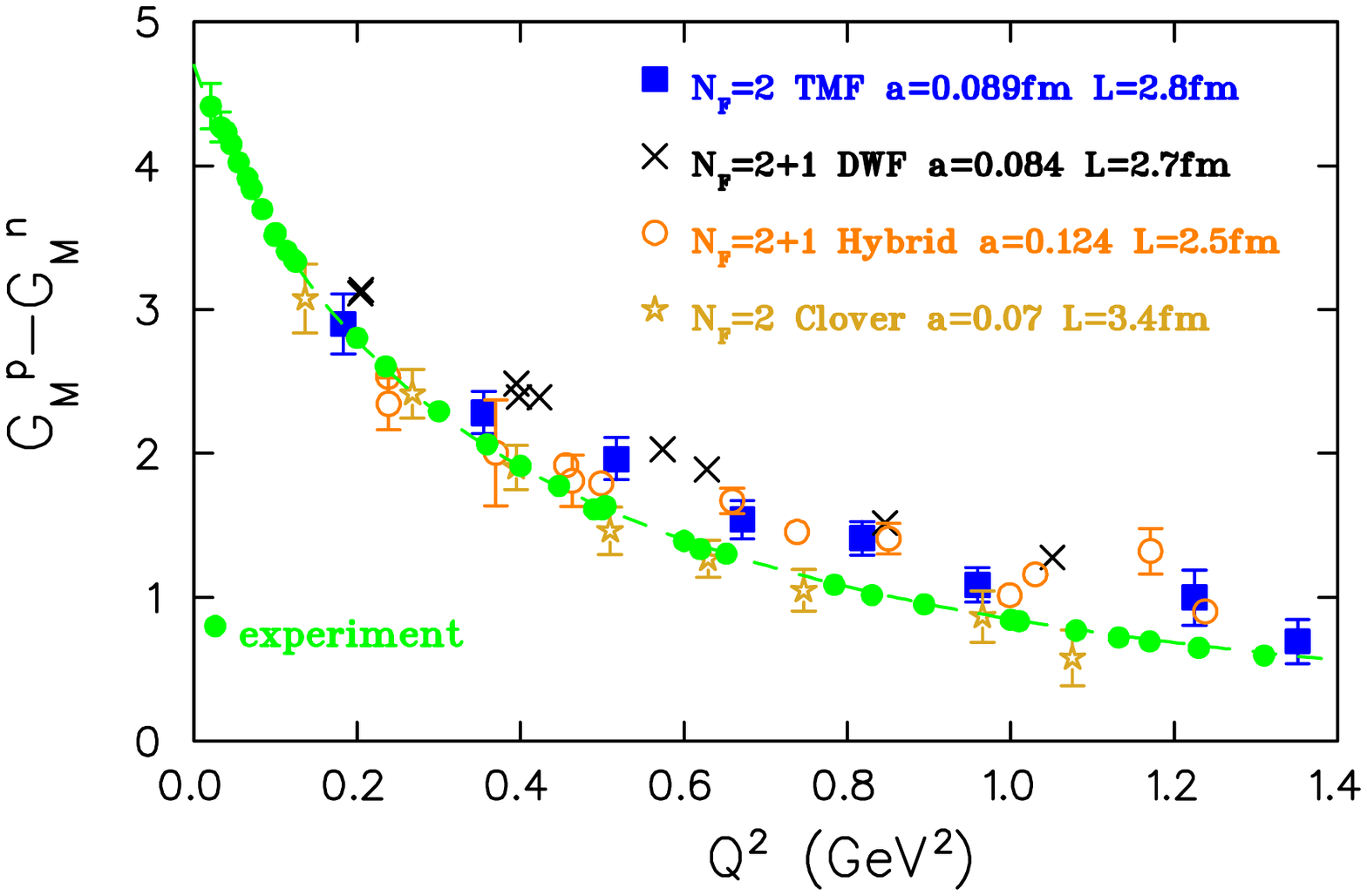}}
   \end{minipage}
\begin{minipage}{0.49\linewidth}
   {\includegraphics[width=1.1\linewidth]{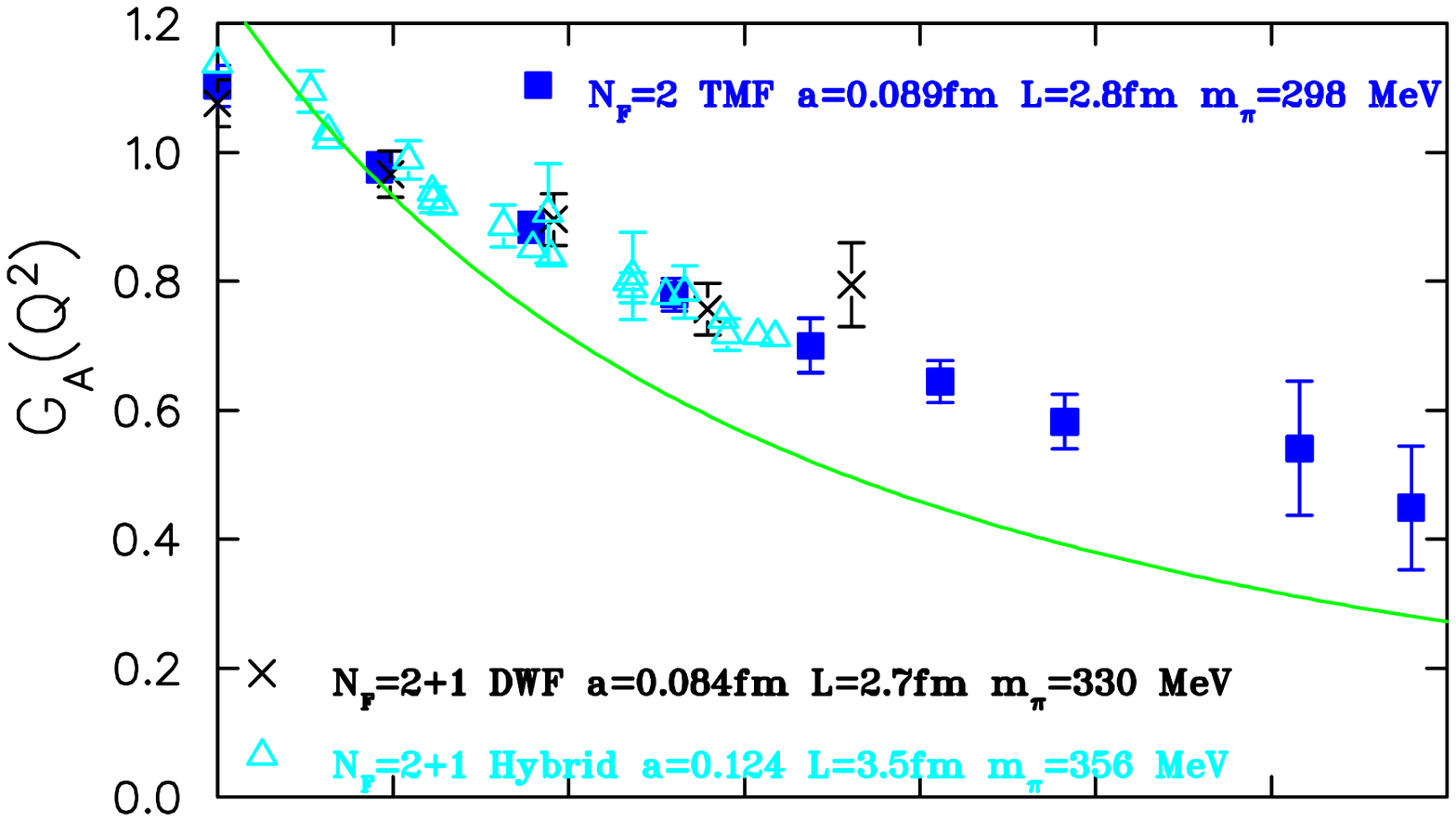}}\vspace*{-6.8cm}
     {\includegraphics[width=1.1\linewidth]{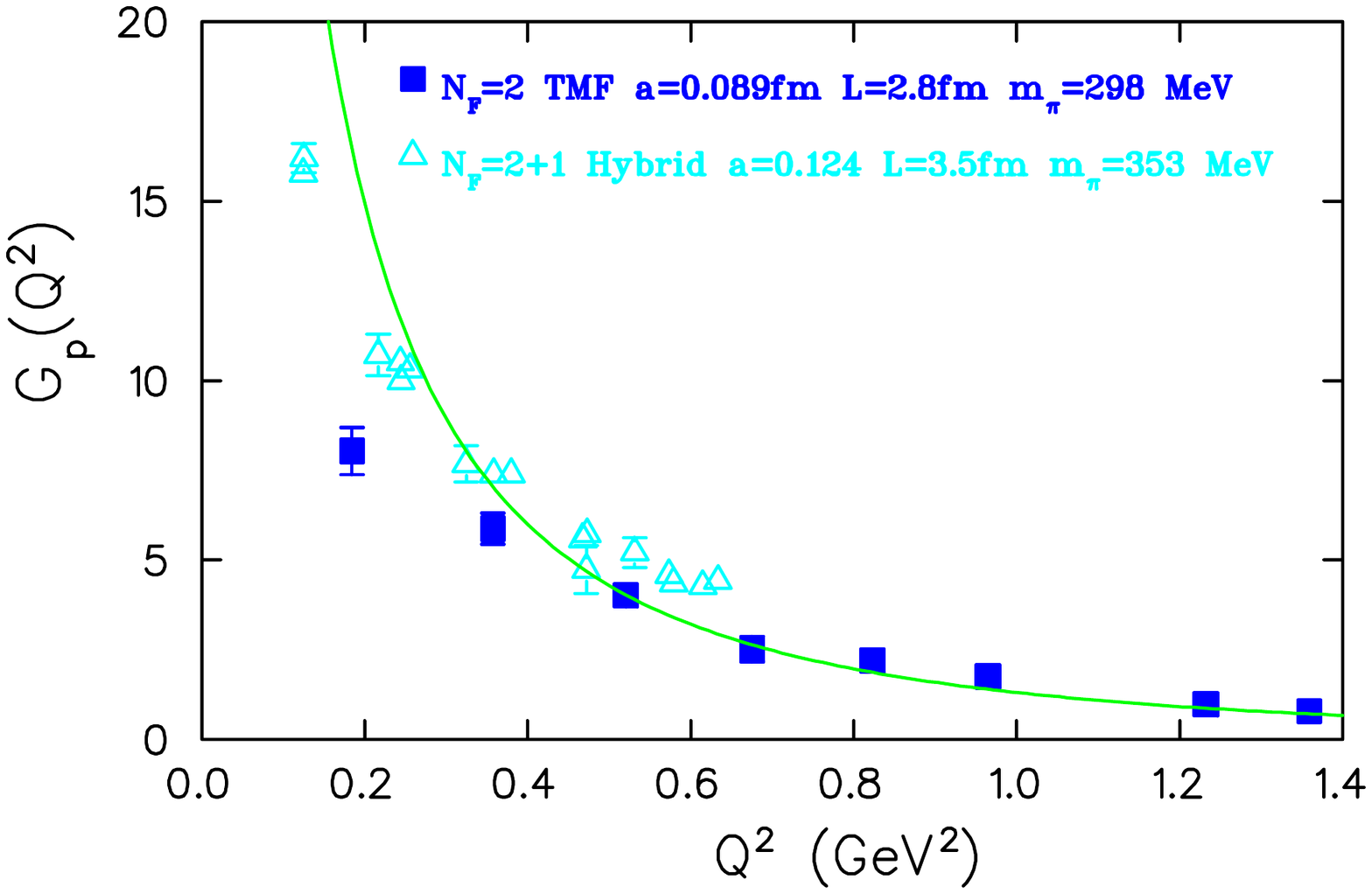}}
   \end{minipage}\vspace*{-2.5cm}
\caption{Left: Isovector electric and magnetic nucleon FFs at $m_\pi\sim 300$~MeV using TMF~\cite{Alexandrou2010}, DWF~\cite{FF-Syritsyn}, hybrid~\cite{FF-Bratt}  and Clover~\cite{Wittig}. Experimental data are shown with the filled circles 
accompanied with Kelly's parametrization shown with the dashed line.  
Right: Axial nucleon FFs. The solid line is a dipole fit to experimental data for $G_A(Q^2)$ combined with pion pole dominance to
get the solid curve shown for $G_p(Q^2)$.}
\label{fig:nucleon EM}
\end{figure*}

We take the continuum limit of TMF results by ﬁtting to a constant, after checking that a linear ﬁt at two values of the pion mass yields consistent results with the constant ﬁt. Volume corrections are taken into account following
  Ref.~\cite{Khan}. The 
volume corrected continuum results are shown in Fig.~\ref{fig:gA}. Chiral extrapolation
using HB$\chi$PT 
with three fit parameters  in the small scale expansion (SSE)
~\cite{SSE} produces a value of $g_A=1.12(8)$ at the
physical point, which is lower than the  experimental value
 by about a standard deviation. The large
error band is due to the large correlations between the $\Delta$ axial
charge $g_{\Delta \Delta}$ and the counter-term involved in the chiral expansion.
 A lattice determination of $g_{\Delta \Delta}$~\cite{Alexandrou:2010tj} will therefore allow a more controlled chiral extrapolation.

In Fig.~\ref{fig:nucleon EM} we show recent lattice results on the EM isovector FFs $G_E(Q^2)$ and $G_M(Q^2)$ at $m_\pi\sim 300$~MeV, where we see
a nice agreement  for $G_E(Q^2)$
but a clear disagreement with experiment, with lattice data showing a weaker
 $Q^2$-dependence. 
For $G_M(Q^2)$ there
is some spread in the results that needs to be investigated.
 For the axial $G_A(Q^2)$
FF there is good agreement among TMF, DWF and hybrid results. Like in the
case of the EM form factors, the $Q^2$-dependence of $G_A(Q^2)$ is milder than what is observed  experimentally. TMF results on $G_p(Q^2)$ at $m_\pi \sim 300$~MeV on a lattice with spatial $L=2.8$~fm and results using a hybrid action of DWF on staggered sea at $m_\pi\sim 350$~MeV
and $L=3.5$~fm show discrepancies 
 at low $Q^2$, that may be due to the smaller volume in our calculations.

 \vspace*{-0.3cm}

\section{Nucleon moments}

\begin{figure}[h]
\begin{minipage}{0.48\linewidth}
{\includegraphics[width=\linewidth]{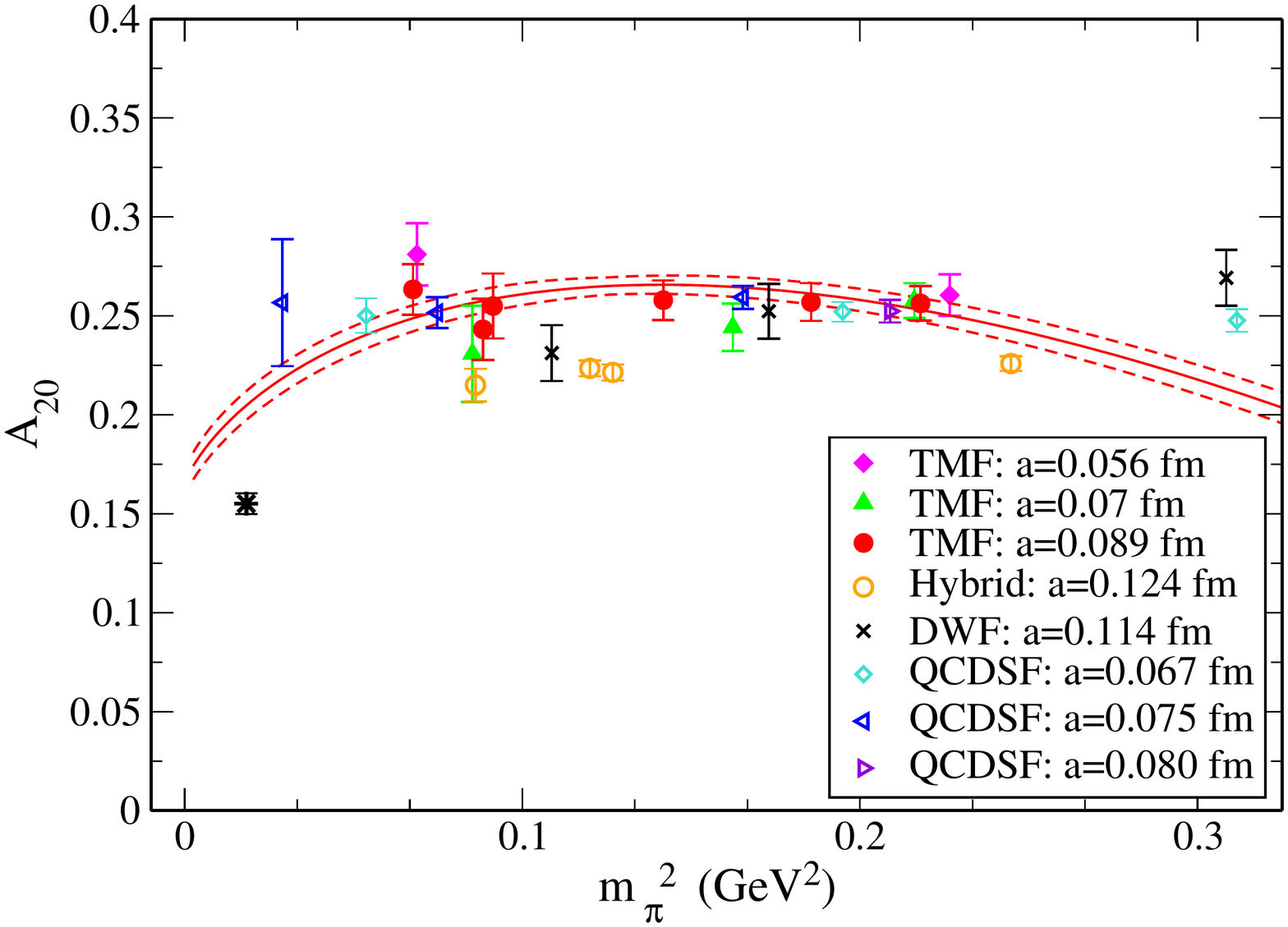}}
\end{minipage}\hfill
\begin{minipage}{0.48\linewidth}
{\includegraphics[width=\linewidth]{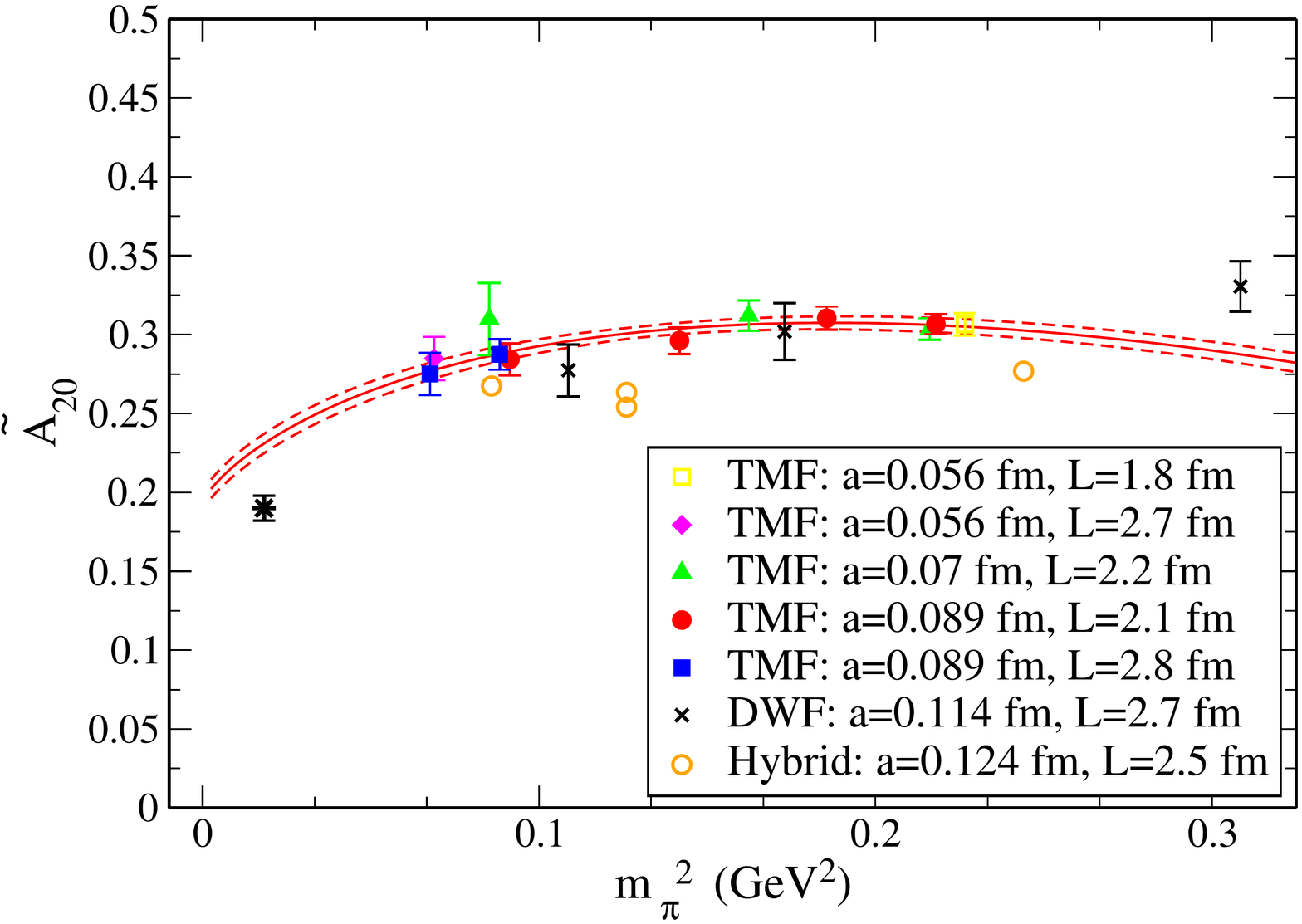}}
\end{minipage}
\vspace*{-0.5cm}
\caption{Recent results on $\langle x \rangle _{u-d}$ and $\langle x \rangle _{\Delta u-\Delta d}$. The fit is done using HB$\chi$PT~\cite{Thomas}.}
 \vspace*{-0.3cm}
\label{fig:GPDs}
\end{figure}

In Figs.~\ref{fig:GPDs} we compare recent results from ETMC~\cite{Alexandrou2010}, RBC-UKQCD~\cite{Aoki}, QCDSF~\cite{GPD-Zanotti} and LHPC~\cite{FF-Bratt} on
the spin-independent and helicity quark distributions.
All collaborations except LHPC use non-perturbative renormalization constants. The ETMC has, in addition, subtracted ${\cal O}(a^2)$ terms perturbatively~\cite{Z-Alexandrou}.
There is a spread among lattice results with results obtained with the hybrid action being lower than those from ETMC and QCDSF.  Using HB$\chi$PT~\cite{Thomas} we extrapolate results on 
$A_{20}$ and $\tilde{A}_{20}$ to the physical point as shown by the curves in
Fig.~\ref{fig:GPDs}. Our estimates for both
 $A_{20}=\langle x \rangle _{u-d}$and $\tilde{A}_{20}=\langle x \rangle _{\Delta u-\Delta d}$ are considerably higher than the experimental values.
The very recent result by QCDSF~\cite{GPD-Zanotti} at $m_\pi\sim 170$~MeV 
 remains higher than experiment and  highlights the need to understand such deviations.
 \vspace*{-0.3cm}

\vspace*{-0.3cm}

\begin{thebibliography}{99}
\bibitem{Alexandrou2010} C. Alexandrou, PoS {\bf Lattice 2010}, 001 (2010), arXiv:1011.3660; C. Alexandrou {\it et al.}, (ETMC) PoS {\bf LATTICE2008}, 139 (2008); C. Alexandrou {\it et al.} (ETMC), PoS {\bf LAT2009} 145 (2009).
\bibitem{Z-Alexandrou}
  C.~Alexandrou, M.~Constantinou, T.~Korzec, H.~Panagopoulos and F.~Stylianou,
  arXiv:1006.1920.
\bibitem{Khan} A. Ali Khan, {\it et al.} (QCDSF), PRD {\bf 74}, 094508 (2006).
\bibitem{SSE} T. R. Hemmert, M. Procure and W. Weise, Phys. Rev. D {\bf 68}, 075009 (2003).
\bibitem{FF-Yamazaki} T. Yamazaki {\it et al.} (RBC-UKQCD), PRD 79, 14505 (2009). 
\bibitem{FF-Bratt}   J. D. Bratt {\it et al.} (LHPC), arXiv:1001.3620.
\bibitem{FF-Syritsyn} S. N. Syritsyn{\it et al.} (LHPC), Phys. Rev. D {\bf 81}, 034507 (2010).
\bibitem{Wittig}
  S.~Capitani, M.~Della Morte, B.~Knippschild and H.~Wittig,
  arXiv:1011.1358.
\bibitem{Alexandrou:2010tj}
  C.~Alexandrou {\it et al.},
  arXiv:1011.0411 [hep-lat].
\bibitem{Aoki} Y. Aoki {\it et al.}, (RBC-UKQCD), Phys. Rev. D {\bf 79}, 114505 (2009).
\bibitem{GPD-Zanotti} J. Zanotti (QCDSF), private communication.
\bibitem{Thomas}  D. Arndt, M. Savage, Nucl. Phys. A 697, 429 (2002);  W.~Detmold, W~Melnitchouk, A.~Thomas, Phys. Rev. D 66, 054501 (2002).


\end{thebibliography}
\end{document}